\pdfoutput=1
\documentclass{article}
\usepackage{spconf,amsmath,graphicx,hyperref}
\usepackage{xspace}

\usepackage[acronym,shortcuts]{glossaries}
\glsdisablehyper    %

\usepackage{amsmath}
\usepackage{amssymb}
\usepackage{bbm}
\usepackage{physics}
\usepackage{siunitx}
\usepackage{trfsigns}
\usepackage{pifont}
\usepackage{etoolbox}
\usepackage{multirow}
\usepackage{csquotes}
\usepackage[shortlabels]{enumitem}  %
\usepackage{cite}   %
\usepackage{hyperref}
\usepackage[capitalize]{cleveref}
\usepackage{xifthen}
\usepackage{balance} %
\usepackage{booktabs}
\usepackage{listings}

\usepackage{makecell}

\usepackage{url}

\let\oldcite\cite
\renewcommand{\cite}[1]{%
\ifthenelse{\isempty{#1}}%
{\inred{[cite!]}}%
{\oldcite{#1}}%
}

\newcolumntype{H}{>{\setbox0=\hbox\bgroup}c<{\egroup}@{}}   %

\newcommand*{\thl}{\fontseries{b}\selectfont}
\newcommand{\diarizen}{DiariZen\xspace}
\robustify\thl
\robustify\fontseries

\newcommand{\inred}[1]{\textcolor{red}{#1}}

\newcommand\defm[2]{\expandafter\newcommand{#1}{\ensuremath{#2}}} %

\defm\Loss{\mathcal{L}}    %
\defm\hwavlm{\vect h_{\text{WavLM}}}  %
\defm\hspat{\vect h_{\text{spat}}}  %

\newcommand{\vect}[1]{\ensuremath{\boldsymbol{\mathbf{#1}}}}

\usepackage{tikz}
\usepackage{pgfplots}
\pgfplotsset{compat=1.16}
\usepgfplotslibrary{groupplots}
\usetikzlibrary{positioning,arrows,matrix,fit,calc,patterns,chains,scopes,shapes.multipart,decorations,decorations.markings,backgrounds,shadows.blur}

\definecolor{palette-1}{HTML}{1f77b4}
\definecolor{palette-2}{HTML}{ff7f0e}
\definecolor{palette-3}{HTML}{2ca02c}
\definecolor{palette-4}{HTML}{d62728}
\definecolor{palette-5}{HTML}{9467bd}
\definecolor{palette-6}{HTML}{8c564b}
\definecolor{palette-7}{HTML}{e377c2}
\definecolor{palette-8}{HTML}{7f7f7f}
\definecolor{palette-9}{HTML}{bcdb22}
\definecolor{palette-10}{HTML}{17becf}

\tikzset{
    line/.style={draw,black,thick,rounded corners=1mm,line cap=round},
    noshortarrow/.style={line,->},
    arrow/.style={noshortarrow,shorten >=.3mm},
    doublearrow/.style={arrow,<->, shorten <=.3mm},
    smallbox/.style={draw,black,thick,rounded corners=3,fill=white,align=center},
    box/.style={draw,black,thick,minimum height=3em,text depth=0.25ex,rounded corners=3,fill=white,align=center},
    nopadding/.style={minimum height=0,inner sep=1mm},
    signalbox/.style={draw,black,thin,rounded corners=1mm,minimum width=7mm, minimum height=4mm,inner sep=0},
    pbox/.style={box,fill=black!10},
    backgroundbox/.style={inner xsep=3mm, inner ysep=1mm, draw, dashed, rounded corners,fill=orange!10},
    branch/.style={inner sep=0.3mm,circle,fill=black},
    operator/.style={draw,circle,black,rounded corners,inner sep=0,fill=white},
    vertex/.style={draw,ultra thin,circle,black,rounded corners,inner sep=0.6mm,fill=gray,fill opacity=0.5},
    edge/.style={line,very thick,line cap=butt},
    pattern1/.style={pattern=north west lines,pattern color=palette-1},
    pattern2/.style={pattern=north east lines,pattern color=palette-2},
    pattern3/.style={pattern=crosshatch,pattern color=palette-3},
    buswidth/.style={path picture={\draw[black,-] (path picture bounding box.south west) -- (path picture bounding box.north east);}}
}

\pgfdeclarelayer{foreground}
\pgfdeclarelayer{background}
\pgfsetlayers{background,main,foreground}

\title{On the Role of Spatial Features in\\ Foundation-Model-Based Speaker Diarization}
\name{
    \begin{tabular}{c}
        Marc Deegen$^{1}$, Tobias Gburrek$^{1}$, Tobias Cord-Landwehr$^{1}$, \\
        Thilo von Neumann$^{1}$, Jiangyu Han$^{2}$, Lukáš Burget$^{2}$, Reinhold Haeb-Umbach$^{1}$
    \end{tabular}
}
\address{
    $^{1}$ Paderborn University, Communications Engineering Department, Germany \\
    $^{2}$ Brno University of Technology, Speech@FIT, Czechia \\
    \small\ttfamily
    \{deegen,gburrek,cord,vonneumann,haeb\}@nt.upb.de \\
    \small\ttfamily
    \{ihan, burget\}@fit.vut.cz
}

\begin{document}
\makeatletter
\renewcommand\section{\@startsection {section}{1}{\z@}%
                                   {-3.5ex \@plus -1ex \@minus -.2ex}%
                                   {1.2ex \@plus.2ex}%
                                   {\normalfont\Large\bfseries}}
\renewcommand\subsection{\@startsection{subsection}{2}{\z@}%
                                     {-3.25ex\@plus -1ex \@minus -.2ex}%
                                     {1ex \@plus .2ex}%
                                     {\normalfont\large\bfseries}}
\renewcommand\subsubsection{\@startsection{subsubsection}{3}{\z@}%
                                     {-3.25ex\@plus -1ex \@minus -.2ex}%
                                     {0.7ex \@plus .2ex}%
                                     {\normalfont\normalsize\bfseries}}
\makeatother
\setlength{\abovedisplayskip}{5pt}
\setlength{\belowdisplayskip}{5pt}
\setlength{\textfloatsep}{7pt plus 2.0pt minus 0.0pt}
\setlength{\floatsep}{5pt plus 0.0pt minus 0.0pt}
\ninept
\maketitle
\begin{abstract}
Recent advances in speaker diarization exploit large pretrained foundation models, such as WavLM, to achieve state-of-the-art performance on multiple datasets. Systems like DiariZen leverage these rich single-channel representations, but are limited to single-channel audio, preventing the use of spatial cues available in multi-channel recordings. This work analyzes the impact of incorporating spatial information into a state-of-the-art single-channel diarization system by evaluating several strategies for conditioning the model on multi-channel spatial features. Experiments on meeting-style datasets indicate that spatial information can improve diarization performance, but the overall improvement is smaller than expected for the proposed system, suggesting that the features aggregated over all WavLM layers already capture much of the information needed for accurate speaker discrimination, also in overlapping speech regions. These findings provide insight into the potential and limitations of using spatial cues to enhance foundation model-based diarization.

\end{abstract}
\section{Introduction}
\label{sec:intro}
Speaker diarization is a fundamental component in many speech processing systems, such as meeting transcription and multi-speaker Automatic Speech Recognition (ASR)~\cite{Park2021ARO,22_horiguchi_eda_eend, 20_medennikov_tsvad}. 
It answers the question of \enquote{who spoke when}, predicting the temporal activity of each speaker in an input recording.
This diarization information can be used to enhance the performance of subsequent downstream tasks.

Different paradigms to diarization exist. Conventional modular diarization systems rely on extracting and clustering speaker representations, such as x-vectors~\cite{8461375}. End-to-End Neural Diarization (EEND) approaches directly predict frame-wise speaker activity from the input audio~\cite{fujita19_interspeech, 9003959}.
A hybrid approach between these two paradigms is the End-to-End Neural Diarization with Vector Clustering (EEND-VC) framework~\cite{kinoshita21_interspeech}, which performs EEND locally on short segments of the recording and subsequently stitches the segment-level predictions by clustering extracted speaker embeddings across segments.

The introduction of large pretrained foundation models like WavLM~\cite{Chen2022_WavLM} has significantly improved speaker diarization performance. By learning from large amounts of unlabeled data, WavLM provides powerful speech representations that effectively reduce the reliance on task-specific training datasets.
The DiariZen~\cite{han2025leveraging, han2025efficient} system makes use of this approach by integrating WavLM-derived features into a Conformer-based~\cite{gulati20_interspeech} EEND model within an EEND-VC framework, achieving state-of-the-art diarization performance. 

However, since most foundation models are pretrained exclusively on single-channel audio, systems that rely on these representations are 
unable to leverage the spatial information present in multi-channel recordings.
In contrast to that, there are approaches that explicitly exploit spatial information for diarization, e.g. in the form of Time Difference of Arrival (TDOA)~\cite{gburrek_asilomar, cordlandwehr25_interspeech, shota_mceend} or Direction of Arrival (DOA)~\cite{4538680} estimates of the received speech.
Spatial information has proven especially beneficial for regions of overlapping speech, where purely spectral systems often struggle, while spatial methods can more effectively separate and attribute concurrent speakers if they are active from different positions in space~\cite{wang2022spatial, cordlandwehr25_interspeech, 4538680, gburrek_asilomar}.
However, spatial systems are typically trained on much smaller datasets 
compared to single-channel systems, which can benefit from large amounts of pretraining data~\cite{Chen2022_WavLM}. 

To take advantage of multi-channel input in single-channel diarization systems, DOVER-Lap can be employed, which combines the output from individual channels to a joint diarization hypothesis~\cite{Raj2021Doverlap}. A computationally less demanding, however even more effective approach was presented in~\cite{han2025spatial}, where inter-channel communication modules were integrated into the early layers of the WavLM feature extraction, thus making WavLM multi-channel aware.

In this contribution, we follow an alternative approach. We develop an auxiliary network tasked to extract spatial information from the multi-channel input, and integrate its output with the single-channel WavLM features.
This integration aims to enable the system to leverage spatial information in addition to the semantic and acoustic representations captured by the WavLM features.

To this end, multiple options of integrating a spatial auxiliary network into the DiariZen diarization pipeline are 
analyzed on their applicability to support the diarization performance.
Here, both a direct incorporation of embeddings derived from spatial features using a neural network and the fusion with a pretrained spatial diarization module are evaluated and analyzed on several meeting-style datasets.
Furthermore, the auxiliary network is designed to be agnostic to both the number of input channels and the microphone array geometry, so as not to restrict the original system to a specific microphone array.

The remainder of this paper is organized as follows.
Section~\ref{cha:sys} provides an overview of the DiariZen framework and details on the proposed integration of spatial features using an auxiliary multi-channel network.
Section~\ref{cha:exp} describes the datasets and experimental setup, followed by a presentation and analysis of the diarization performance in terms of Diarization Error Rate (DER), and conclusions are drawn in \cref{sec:conclusions}.

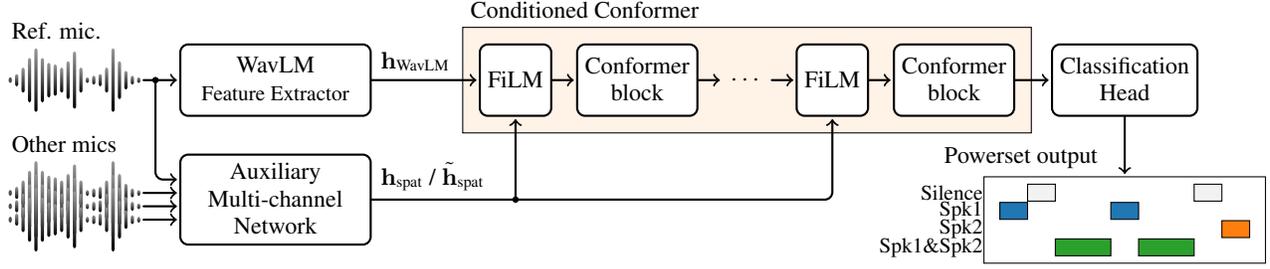
\begin{figure*}[bt]
    \centering
    \begin{tikzpicture}[
    wav/.style={
        minimum width=13ex,
        minimum height=6ex,
        path picture={
            \foreach \size [count=\i] in {0.1, 0.2, 0.4, 0.7, 1, 0.7, 0.55, 0.5, 0.4, 0.6, 0.9, 0.4, 0.1, 0.15, 0.2, 0.4, 1, 0.6, 0.3, 0.15, 0.1} {
                \def\x{\i*.6ex}
                \def\y{\size*3ex}
                \path[top color=black!40,bottom color=black,rounded corners=1] ($(path picture bounding box.west) + (\x-.15ex,-\y)$) rectangle ($(path picture bounding box.west) + (\x+.15ex,\y)$);
            }
        }
    }
]
    \node[box,minimum width=8em] (wavlm) {WavLM\\\footnotesize Feature Extractor};
    \node[box,right=4.5em of wavlm] (film1) {FiLM};
    \node[box,right=1em of film1] (conf1) {Conformer\\block};
    \node[right=1em of conf1] (dots) {$\cdots$};
    \node[box,right=1em of dots] (film2) {FiLM};
    \node[box,right=1em of film2] (conf2) {Conformer\\block};

    \node[box,right=1.5em of conf2] (classification) {Classification\\Head};

    \draw[arrow] (wavlm) -- node[pos=0,anchor=south west] {$\hwavlm$} (film1);
    \draw[arrow] (film1) -- (conf1);
    \draw[arrow] (conf1) -- (dots);
    \draw[arrow] (dots) -- (film2);
    \draw[arrow] (film2) -- (conf2);
    \draw[arrow] (conf2) -- (classification);

    \node[matrix of nodes,below=2.5em of classification,nodes in empty cells,column sep=.5em,draw] (m) {
         &&&&&&&&& \\[0.1em]
         &&&&&&&&& \\[0.1em]
         &&&&&&&&& \\[0.1em]
         &&&&&&&&& \\
    };
    \draw[fill=palette-1] (m-2-1.north) rectangle (m-2-2.south);
    \draw[fill=gray!10] (m-1-2.north) rectangle (m-1-3.south);
    \draw[fill=palette-3] (m-4-3.north) rectangle (m-4-5.south);
    \draw[fill=palette-1] (m-2-5.north) rectangle (m-2-6.south);
    \draw[fill=palette-3] (m-4-6.north) rectangle (m-4-8.south);
    \draw[fill=gray!10] (m-1-8.north) rectangle (m-1-9.south);
    \draw[fill=palette-2] (m-3-9.north) rectangle (m-3-10.south);

    \node[left]  at (m-1-1.west) {\footnotesize Silence};
    \node[left]  at (m-2-1.west) {\footnotesize Spk1};
    \node[left]  at (m-3-1.west) {\footnotesize Spk2};
    \node[left]  at (m-4-1.west) {\footnotesize Spk1\&Spk2};

    \node[anchor=south west,xshift=-2em] at (m.north west) {Powerset output};

    \draw[arrow] (classification) -- (m);

    \node[left=1.5em of wavlm, wav] (wav1) {};
    \draw[arrow](wav1) -- (wavlm);
    \node[anchor=south west] at (wav1.north west) {Ref. mic.};

    \def\wavoffset{1.25ex}
    \node[wav,below=2em of wav1] (wav2) {};
    \node[wav,yshift=-\wavoffset] (wav3) at (wav2) {};
    \node[wav,yshift=-\wavoffset] (wav4) at (wav3) {};
    \node[anchor=south west] at (wav2.north west) {Other mics};

    \coordinate (wavcenter) at ($(wav2)!.5!(wav3)$);
    \node[box,minimum width=8em] at (wavlm|-wavcenter) (encoder) {Auxiliary\\Multi-channel\\Network};
    \draw[arrow] (wav1.east) ++ (.5em,0) node[branch]{} |- ($(encoder.west) + (0,1.5*\wavoffset)$);
    \draw[arrow] (wav2.east) -- (wav2-|encoder.west);
    \draw[arrow] (wav3.east) -- (wav3-|encoder.west);
    \draw[arrow] (wav4.east) -- (wav4-|encoder.west);

    \draw[arrow] (encoder) -| node[above,pos=0,anchor=south west] {$\hspat$ / $\tilde{\mathbf{h}}_{\text{spat}}$} (film2);
    \draw[arrow] (encoder-|film1) node[branch]{} -- (film1);

    \begin{pgfonlayer}{background}
        \node[draw,fit={(film1)(conf2)},fill=palette-2!10,inner sep=1.5ex] (conformer) {};
    \end{pgfonlayer}
    \node[anchor=south west] at (conformer.north west) {Conditioned Conformer};
\end{tikzpicture}
    \vspace{-1em}
   \caption{Overview of the spatially supported DiariZen architecture. First, the single-channel WavLM features are extracted and then combined with spatial cues, using FiLM layers. The spatial cues are extracted by an auxiliary multi-channel network, which takes spatial features consisting of IPDs and magnitude as input.}
   \vspace{-1.5em}
    \label{fig:system-overview}
\end{figure*}

\section{Spatially Supported DiariZen}
\label{cha:sys}

The analysis in this work is based on the single-channel multi-speaker diarization framework DiariZen~\cite{han2025leveraging}.
To investigate the impact of spatial information extracted from multi-channel signals on the diarization, DiariZen is extended with a spatial feature extraction module, as illustrated in \cref{fig:system-overview}.
Here, a compact microphone array setup is assumed, providing spatial cues such as inter-channel phase differences (IPDs)~\cite{song21_interspeech} and magnitude information, which are closely related to the information used for diarization in TDOA- and DOA-based approaches.
These features are combined with the WavLM features from the DiariZen framework
to analyze whether spatial information can further enhance diarization performance. %

\subsection{DiariZen}

\diarizen{} follows the EEND-VC~\cite{kinoshita21_interspeech,kinoshita21_icassp} framework, where the EEND module uses a WavLM~\cite{Chen2022_WavLM} feature extractor, fine-tuned to the diarization scenario.
In the EEND-VC framework, the input audio is divided into short overlapping segments.
On each segment independently, an EEND~\cite{fujita19_interspeech, 9003959} model first estimates frame-level speaker activities, producing local diarization outputs. Since each segment is processed independently, speaker identities are not consistent across segments, requiring an additional alignment and merging stage to resolve the speaker label ambiguity. 
This is achieved by a subsequent Vector Clustering (VC) process:
Speaker embeddings are extracted for each locally detected speaker from non-overlapping speech regions and clustered across the full recording using agglomerative hierarchical clustering (AHC), or Variational Bayes HMM clustering with x-vectors (VBx)~\cite{fede_vbx},
with the constraint that embeddings originating from the same segment cannot be merged.

\begin{figure}[b]
    \centering
    \begin{tikzpicture}[
    wav/.style={
        minimum width=8ex,
        minimum height=4ex,
        path picture={
            \foreach \size [count=\i] in {0.1, 0.2, 0.4, 0.7, 1, 0.7, 0.55, 0.5, 0.4, 0.6, 0.9, 0.4, 0.1, 0.15, 0.2, 0.4, 1, 0.6, 0.3, 0.15, 0.1} {
                \def\x{\i*.6ex}
                \def\y{\size*3ex}
                \path[top color=black!40,bottom color=black,rounded corners=1] ($(path picture bounding box.west) + (\x-.15ex,-\y)$) rectangle ($(path picture bounding box.west) + (\x+.15ex,\y)$);
            }
        }
    }
]
    \node[box,] (wavlm) {WavLM};
    \node[box,right=1em of wavlm] (sum) {Weighted\\Sum};
    \node[box,right=1em of sum] (linear) {Linear\\+ LN};
    \node[box,right=1em of linear] (conf1) {Conformer};
    \node[right=1em of conf1] (end) {};

    \draw[arrow] (wavlm) -- (sum);
    \draw[arrow] (sum) -- (linear);
    \draw[arrow] (linear) -- (conf1);
    \draw[arrow] (conf1) -- (end);

    \node[left=1.5em of wavlm, wav] (wav1) {};
    \draw[arrow](wav1) -- (wavlm);
    \node[anchor=south west] at (wav1.north west) {Ref. mic.};

    \begin{pgfonlayer}{background}
        \node[draw,fit={(wavlm)(linear)},fill=palette-2!10,inner sep=1ex] (conformer) {};
    \end{pgfonlayer}
    \node[anchor=south west] at (conformer.north west) {WavLM Feature Extractor};
\end{tikzpicture}
    \vspace{-2em} 
    \caption{Illustration of the local EEND module from the DiariZen framework (adapted from \protect\cite{han2025leveraging}).}
    \label{fig:diarizen}
\end{figure}
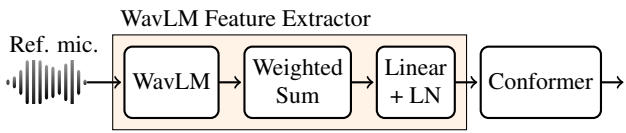
This work focuses on the local EEND module of the DiariZen framework visualized in \cref{fig:diarizen}.
DiariZen extracts WavLM features, obtained by combining the outputs from all WavLM layers using a learnable weighted sum.
The aggregated WavLM features are projected through a linear layer followed by layer normalization, and then passed to a Conformer with a classification head trained using powerset classification to predict the diarization output~\cite{plaquet23_interspeech}.
In powerset classification, all possible combinations of active speakers, including the silence class, single active speakers, and overlapped speakers, are represented as distinct target classes. 
This approach effectively converts the multi-label speaker activity detection problem into a single-label multi-class classification problem, which can be optimized using a cross-entropy loss.

\subsection{Auxiliary multi-channel network}
\label{aux_model}

Spatial information is gathered by computing the IPD features $\text{IPD}_m$ for the $m$-th of all $M$ non-redundant microphone pairs. 
To address the inherent phase discontinuity, sine and cosine transformations of the phase are applied in order to yield a continuous representation of the phase~\cite{cosIPD_2018}.
Those, as well as the magnitude spectrogram $y_{\text{ref}}$ of the first microphone channel, constitute the spatial input features of the auxiliary multi-channel network. %

\subsubsection{Spatial encoder}
\label{cha:enc}

\begin{figure}[bt]
    \centering
    \begin{tikzpicture}
    \node (abs) {$|y_{\text{ref}}|$};
    \node[below=1.5em of abs.east, left] (cos1) {$\cos (\text{IPD}_1)$};
    \node[below=1.5em of cos1.east,left] (sin1) {$\sin (\text{IPD}_1)$};
    \node[below=1.5em of sin1.east,left] (cos2) {$\cos (\text{IPD}_2)$};
    \node[below=3em of cos2.east,left] (sin2) {$\sin (\text{IPD}_{M})$};
    \node[yshift=.5ex] at ($(cos2)!0.5!(sin2)$) (dots) {\strut$\vdots$};

    \node[smallbox,right=1.25em of abs,align=center] (SAabs) {Self-Attention};
    \node[smallbox,right=1.25em of cos1,align=center] (SAcos1) {Self-Attention};
    \node[smallbox,right=1.25em of sin1,align=center] (SAsin1) {Self-Attention};
    \node[smallbox,right=1.25em of cos2,align=center] (SAcos2) {Self-Attention};
    \node[smallbox,right=1.25em of sin2,align=center] (SAsin2) {Self-Attention};

    \draw[arrow] (abs) -- (SAabs);
    \draw[arrow] (cos1) -- (SAcos1);
    \draw[arrow] (sin1) -- (SAsin1);
    \draw[arrow] (cos2) -- (SAcos2);
    \draw[arrow] (sin2) -- (SAsin2);

    \node[smallbox,fit={(SAabs.north east)(SAsin2.south east)},xshift=2em,minimum width=1.5em] (tac) {};
    \node[rotate=90] at (tac) {TAC};

    \node[right=3.25em of SAabs] (dotsabs) {$\dots$};
    \node[right=3.25em of SAcos1] (dotscos1) {$\dots$};
    \node[right=3.25em of SAsin1] (dotssin1) {$\dots$};
    \node[right=3.25em of SAcos2] (dotscos2) {$\dots$};
    \node[right=3.25em of SAsin2] (dotssin2) {$\dots$}; 

    \node[smallbox,fit={(SAabs.north east)(SAsin2.south east)},xshift=7em,minimum width=1.5em] (avg) {};
    \node[rotate=90] at (avg) {Average};

    \draw[arrow] (SAabs) -- (tac.west|-SAabs);
    \draw[arrow] (SAcos1) -- (tac.west|-SAcos1);
    \draw[arrow] (SAsin1) -- (tac.west|-SAsin1);
    \draw[arrow] (SAcos2) -- (tac.west|-SAcos2);
    \draw[arrow] (SAsin2) -- (tac.west|-SAsin2);

    \draw[arrow] (tac.east|-SAabs) -- (dotsabs) -- (avg.west|-SAabs);
    \draw[arrow] (tac.east|-SAcos1) -- (dotscos1) -- (avg.west|-SAcos1);
    \draw[arrow] (tac.east|-SAsin1) -- (dotssin1) -- (avg.west|-SAsin1);
    \draw[arrow] (tac.east|-SAcos2) -- (dotscos2) -- (avg.west|-SAcos2);
    \draw[arrow] (tac.east|-SAsin2) -- (dotssin2) -- (avg.west|-SAsin2);

    \draw[arrow] (avg.east) -- ++(1.5em,0) node[right] {$\hspat$};

    \begin{pgfonlayer}{background}
        \node[fit={(SAabs)(SAcos1)(SAsin1)(SAcos2)(SAsin2)(tac)},draw,fill=palette-2!10,inner xsep=1.5ex] (block) {};
    \end{pgfonlayer}
    \node[anchor=north west,inner ysep = 0,inner xsep=1pt] at (block.north east) {$\times N$};
\end{tikzpicture}    
    \vspace{-0.5em}
     \caption{
        Spatial encoder architecture to estimate the spatial cues $\vect h_{\text{spatial}}$. 
        $N$ encoder layers, with shared weights self-attention and TAC connections across all transformed input features, are stacked before the output is averaged after the last layer. 
    }
    \label{fig:gcc-encoder}
\end{figure}
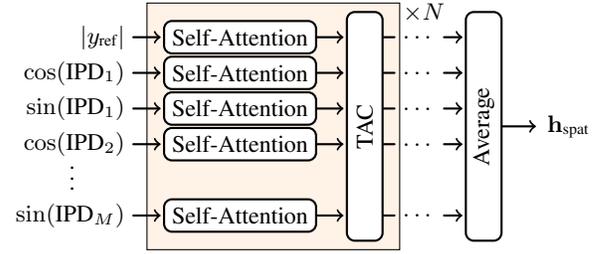

Two variants of the auxiliary network are tested in this work.
The first, referred to as the spatial encoder, is illustrated in \cref{fig:gcc-encoder}. %
At each encoder layer, self-attention with shared weights is applied to all spatial features and inter-channel interactions are facilitated through Transform, Average, and Concatenate (TAC)~\cite{luo2020end} connections after the self-attention.
This architecture enables cross-channel information exchange and transforms the spatial features into an embedding space such that, after the final encoder layer, the representations can be averaged across channels without losing essential spatial information, yielding a single-channel spatial embedding~$\hspat$.
Consequently, the resulting spatial encoder and subsequent modules are agnostic to both the number of input channels and the specific microphone array configuration. 

\subsubsection{Spatial conformer and spatial diarization}
\label{cha:spatial_diarization}

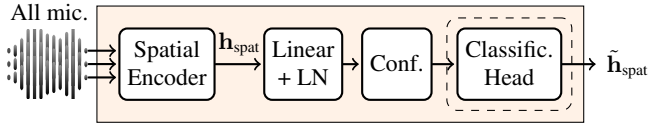
\begin{figure}
    \centering
    \begin{tikzpicture}[
    wav/.style={
        minimum width=8ex,
        minimum height=4ex,
        path picture={
            \foreach \size [count=\i] in {0.1, 0.2, 0.4, 0.7, 1, 0.7, 0.55, 0.5, 0.4, 0.6, 0.9, 0.4, 0.1, 0.15, 0.2, 0.4, 1, 0.6, 0.3, 0.15, 0.1} {
                \def\x{\i*.6ex}
                \def\y{\size*3ex}
                \path[top color=black!40,bottom color=black,rounded corners=1] ($(path picture bounding box.west) + (\x-.15ex,-\y)$) rectangle ($(path picture bounding box.west) + (\x+.15ex,\y)$);
            }
        }
    }
]
    \node[box,] (enc) {Spatial\\Encoder};
    \node[box,right=2em of enc] (linear) {Linear\\+ LN};
    \node[box,right=.75em of linear] (conf) {Conf.};
    \node[box,right=1em of conf] (ch) {Classific.\\Head};
    \node[fit=(ch), draw, dashed, rounded corners, inner sep=0.4em] {};
    \node[right=1.5em of ch] (o) {$\tilde{\mathbf{h}}_{\text{spat}}$};

    \draw[arrow] (enc) -- node[above]{$\hspat$} (linear);
    \draw[arrow] (linear) -- (conf);
    \draw[arrow] (conf) -- (ch);
    \draw[arrow] (ch) -- (o);

    \def\wavoffset{1.25ex}
    \node[wav,above left=-1.9em and 1.25em of enc] (wav2) {};
    \node[wav,yshift=-\wavoffset] (wav3) at (wav2) {};
    \node[wav,yshift=-\wavoffset] (wav4) at (wav3) {};
    \node[anchor=south west] at (wav2.north west) {All mic.};

    \coordinate (wavcenter) at ($(wav2)!.5!(wav3)$);
    \draw[arrow] (wav2.east) -- (wav2-|enc.west);
    \draw[arrow] (wav3.east) -- (wav3-|enc.west);
    \draw[arrow] (wav4.east) -- (wav4-|enc.west);

    \begin{pgfonlayer}{background}
        \node[draw,fit={(enc)(ch)},fill=palette-2!10,inner sep=2ex] (conformer) {};
    \end{pgfonlayer}
\end{tikzpicture}
    \vspace{-2em}
     \caption{Illustration of the spatial conformer and the spatial diarization networks.
     The features extracted from the spatial encoder are used as input to a network structure similar to \diarizen.}
    \label{fig:conformer}
\end{figure}

Alternatively, the spatial encoder is extended by a projection layer, layer normalization, and an additional conformer, resembling the structure of the single-channel \diarizen system in \cref{fig:diarizen}. The goal is to obtain more complex abstractions $\tilde{\mathbf{h}}_{\text{spat}}$ in the spatial auxiliary network that can be used by the \diarizen network. This configuration is referred to as the ``spatial conformer'' in the following and is illustrated in \cref{fig:conformer}.
When further cascaded with a classification layer, it can be trained as an autonomous spatial diarization module, denoted as ``spatial diarization'' configuration, which may also serve as an auxiliary network to provide spatial cues $\tilde{\mathbf{h}}_{\text{spat}}$ to the \diarizen{} system.

\subsubsection{Spatial conditioning of DiariZen}

The output of the auxiliary network, $\hspat$ or $\tilde{\mathbf{h}}_{\text{spat}}$, is integrated into DiariZen as conditioning input to Feature-wise Linear Modulation (FiLM)~\cite{film} layers, as illustrated in \cref{fig:system-overview}. %
First, one FiLM layer is applied before the Conformer, and then another FiLM layer is applied before each Conformer block to ensure that spatial information is consistently available throughout the network.

\section{Experiments}
\label{cha:exp}
For the experiments, a reimplementation of the pruned and finetuned version of DiariZen, introduced in~\cite{han2025efficient}, is used. 
The hyperparameters in the configuration, like segment length and hop size, follow the setup of the DiariZen framework.
The powerset classification used for training the systems assumes a maximum of 2 concurrent speakers per frame.
Throughout all multi-channel experiments, four microphones are used, irrespective of the total number of microphones available on the respective dataset. The microphones were selected such that the spacing between the microphones was maximized, in order to ensure best capture of spatial information.

Note that the focus of this work is on the performance of the local EEND module. Therefore, in the following evaluations, the segment-level EEND outputs are stitched in an oracle manner.
The estimated local speaker activity is compared to the ground-truth activity to associate a ground-truth speaker label with each local speaker.
The assigned oracle speaker labels are used to resolve the permutation ambiguity between segments.
In this way, the performance evaluation can focus on the performance of the local EEND module, which is where the potential advantage of spatial features should become visible.
Also, no collar is used for the DER computation.

For training and evaluation of the systems, the multi-talker, meeting-style, and multi-channel datasets AMI~\cite{carletta2005ami}, AliMeeting~\cite{yu2022m2met}, AISHELL-4~\cite{fu2021aishell}, and NOTSOFAR-1~\cite{vinnikov2024notsofar} are used. 
Since the AISHELL-4 dataset does not provide an official development set, the same development split as used in DiariZen is adopted.
Contrary to experiments in~\cite{han2025spatial}, the CHiME-6~\cite{watanabe20b_chime} dataset is excluded from this analysis. 
Its multi-room recording setup would likely lead to performance improvements driven primarily by differences in recording conditions across rooms rather than by the effective use of spatial cues, making it unsuitable for a fair evaluation of spatial information effects.
Training is performed on the combined training sets of all four datasets.
\cref{tab:dataset_information} shows the number of active speakers, the size of each dataset and of the combined dataset.

\begin{table}[bt]
\setlength{\tabcolsep}{5pt}
\centering
\vspace{-6pt}
\caption{Dataset properties (\#Spk = \#Speakers, \#Hrs = \#Hours).}

\label{tab:dataset_information}
\vspace{6pt}
\begin{tabular}{l|cc|cc|cc}
\hline
\multirow{2}{*}{\textbf{Dataset}} 
 & \multicolumn{2}{c|}{\textbf{Train}} 
 & \multicolumn{2}{c|}{\textbf{Dev}} 
 & \multicolumn{2}{c}{\textbf{Test}} \\
 & \#Spk & \#Hrs & \#Spk & \#Hrs & \#Spk & \#Hrs \\
\hline
AMI & 3–5 & 79.7 & 4 & 9.7 & 3–4 & 9.1 \\
AISHELL-4 & 3–7 & 97.2 & 3–7 & 10.3 & 5–7 & 12.7 \\
AliMeeting & 2–4 & 111.4 & 2–4 & 4.2 & 2–4 & 10.8 \\ %
NOTSOFAR-1 & 4-8 & 39.8 & 4-6 & 13.4 & 3-7 & 16.5 \\
\hline
Combined & 2–8 & 328.1 & 2–7 & 37.6 & 2–7 & 49.1 \\
\hline
\end{tabular}
\end{table}

\subsection{Reference systems}
\label{cha:baselines}
First, the single-channel DiariZen (ID~1) baseline is evaluated in  \cref{tab:comparison}. It achieves a macro DER of \SI{12.2}{\percent} across the four datasets, with \SI{8.7}{\percent} in single-speaker and \SI{20.1}{\percent} in overlapping speech regions.
The purely spatial system ``Spatial Diarization'' (ID 4), shown in \cref{fig:conformer}, serves as a second baseline. Here, only the auxiliary network as described in \cref{cha:spatial_diarization} is employed for diarization.
It is trained with the same powerset loss and diarization objective as the DiariZen baseline and achieves a macro DER of \SI{14.3}{\percent}, with \SI{10.6}{\percent} in single-speaker regions and \SI{23.6}{\percent} in overlapping regions.

While the overall performance is worse than the DiariZen baseline, the results show that the Spatial Diarization system also does not provide the expected improvement in overlapping speech regions, suggesting that the spatial features offer limited additional benefit for handling overlap.
Since WavLM is originally trained with a masked prediction loss~\cite{Chen2022_WavLM}, it is primarily optimized for single-speaker modeling.
We hypothesize that the surprisingly good performance of the WavLM features in overlap regions might be attributed to the learnable weighted sum across all WavLM layers that allows the model to integrate information also from earlier layers, which are closer to the raw waveform and may already contain cues useful for distinguishing overlapping speakers.
Further analysis of the learned features and weights is left for future work.
Nevertheless, the fact that spatial features alone can be used for an effective diarization suggests that combining spatial and spectral information could further enhance overall diarization performance.

Furthermore, as a topline, DiariZen + Oracle \#Spk (ID 9) is evaluated, which incorporates oracle speaker count information per frame and achieves a macro DER of \SI{4.9}{\percent} (ID~8).
In this setup, the oracle speaker count is used as a conditioning signal to the FiLM layer before the Conformer, while no additional FiLM conditioning is applied within the Conformer.

\subsection{Spatially supported \diarizen{}}

\begin{table*}[!tb]
    \centering
    \vspace{-6pt}
    \caption{DER comparison of the proposed spatially supported systems and the single-/multi-channel \diarizen{} systems using oracle clustering, with separate results for overlapping (OV) and single-speaker (Single) regions.}
    \label{tab:comparison}
    \vspace{3pt}
    \begin{tabular}{clcccccc}
        \toprule 
        \textbf{ID} & \textbf{System} & \makecell{\textbf{AMI}\\(OV / Single)} & \makecell{\textbf{AliMeeting}\\(OV / Single)} & \makecell{\textbf{AISHELL-4}\\(OV / Single)} &  \makecell{\textbf{NOTSOFAR-1}\\(OV / Single)}  & \makecell{\textbf{Macro}\\(OV / Single)}\\
        \midrule
        {1} & DiariZen~\cite{han2025leveraging} & 13.1 (21.7 /\hphantom{1} 9.9) & 12.5 (22.2 / 7.0) & \hphantom{1}9.1 (16.4 / 8.3) & 14.2  (19.9 /\hphantom{1} 9.4) & 12.2 (20.1 /\hphantom{1} 8.7) \\
        {2}& {DiariZen-Large Conformer} & 13.2 (21.3 / 10.2) & 12.6 (22.1 / 7.1) & \hphantom{1}9.6 (16.0 / 8.9) & 14.1  (19.7 /\hphantom{1} 9.5)  &  12.9 (19.8 /\hphantom{1} 8.9) \\
        {3}& {Multi-channel DiariZen~\cite{han2025spatial}} &
        12.8 (\hphantom{9}\hspace{0.2em}- \hphantom{9}\hspace{0.2em}/\hphantom{9}\hspace{0.2em} -\hphantom{9}\hspace{0.2em}) &
        12.0 (\hphantom{8}\hspace{0.1em}- \hphantom{8}\hspace{0.1em}/\hphantom{8}\hspace{0.1em} -\hphantom{8}\hspace{0.1em}) & 
        \hphantom{1}\textbf{8.9} (\hphantom{8}\hspace{0.1em}- \hphantom{8}\hspace{0.1em}/\hphantom{8}\hspace{0.1em} -\hphantom{8}\hspace{0.1em}) & 
        14.1 (\hphantom{9}\hspace{0.2em}- \hphantom{9}\hspace{0.2em}/\hphantom{9}\hspace{0.2em} -\hphantom{9}\hspace{0.2em})  & 
        12.0 (\hphantom{9}\hspace{0.2em}- \hphantom{9}\hspace{0.2em}/\hphantom{9}\hspace{0.2em} -\hphantom{9}\hspace{0.2em}) \\        
        \hline
        {4} & Spatial Diarization  & 14.5 (23.9 / 11.1) & 14.0 (25.4 / 7.4) & 10.0 (22.5 / 8.6) & 18.5 (22.5 / 15.1) & 14.3 (23.6 / 10.6)\\
        {5} & \diarizen{} + Spatial Encoder  & 13.5 (22.4 / 10.2)& 12.6 (22.1 / 7.1) & \hphantom{1}9.5 (17.9 / 8.6) & 14.3 (20.2 /\hphantom{1} 9.4) & 12.5 (20.7 /\hphantom{1} 8.8)\\
        
        {6} & \diarizen{} + Spatial Conformer & 13.5 (22.1 / 10.3) & 13.1 (23.0 / 7.3)  & \hphantom{1}9.4 (18.4 / 8.5) & 14.7 (20.4 /\hphantom{1} 9.9)  & 12.7 (21.0 /\hphantom{1} 9.0)  \\
        {7} & \diarizen{} + Spatial Diarization  & 12.5 (20.8 /\hphantom{1} 9.4) & 12.1 (21.4 / 6.7)  &\hphantom{1}\textbf{8.9} (18.5 / 7.8) & 13.5 (19.0 /\hphantom{1} 8.8)  & 11.7 (19.9 /\hphantom{1} 8.2)\\
        {8} & \quad + Joint Finetuning & \textbf{12.2 }(20.5 / \hphantom{1}9.2)  &  \textbf{11.8} (21.2 / 6.3)  &\hphantom{1}\textbf{8.9} (17.4 / 8.0)  & \textbf{13.4} (18.8 /\hphantom{1} 8.8)  & \textbf{11.6} (19.5 /\hphantom{1} 8.1) \\
        \midrule
        \textcolor{gray!90}{9} & \textcolor{gray!90}{DiariZen + Oracle \#Spk} & \textcolor{gray!90}{\hphantom{1}3.6 (10.1 / \hphantom{1}1.1)} & \textcolor{gray!90}{\hphantom{1}6.0 (14.6 / 1.1)} & \textcolor{gray!90}{\hphantom{1}1.6 (\hphantom{1}4.8 / 1.2)} & \textcolor{gray!90}{\hphantom{1}8.2 (14.7 / \hphantom{1}2.8)} & \textcolor{gray!90}{\hphantom{1}4.9 (11.1 / \hphantom{1}1.6)}\\
        \bottomrule
    \end{tabular}
    \vspace{-1.5em}
\end{table*}

To evaluate this integration of spatial features within the DiariZen framework, the different auxiliary multi-channel networks from \cref{aux_model} are evaluated.
First,  the spatial encoder auxiliary network, illustrated in \cref{fig:gcc-encoder} and described in \cref{cha:enc}, is employed.
During training, the spatial encoder is randomly initialized and jointly trained with the pretrained WavLM and Conformer modules from the pruned DiariZen framework.
However, this DiariZen + Spatial Encoder (ID 5) system achieves a performance comparable to the DiariZen baseline with \SI{12.5}{\percent} macro DER, indicating that the encoded spatial features do not provide a measurable benefit in this configuration.

Given that the spatial encoder is relatively lightweight compared to the other components, a larger model variant is explored in the DiariZen + Spatial Conformer (ID 6) system. %
Here, the auxiliary network from \cref{cha:spatial_diarization}, as shown in \cref{fig:conformer}, without the classification head and without diarization pretraining, is evaluated.
Despite the increased model capacity, the system achieves only a macro DER of \SI{12.7}{\percent}, also not improving over the DiariZen baseline.

Then, the DiariZen + Spatial Diarization (ID 7) configuration is evaluated, integrating the pretrained spatial diarization pipeline from \cref{cha:baselines} as the auxiliary multi-channel network.
This setup extends the Spatial Conformer with a classification head and, more importantly, leverages pretraining on a diarization objective, aiming to provide spatial cues that are more structured and discriminative with respect to speaker activity.
Here, the spatial diarization pipeline remains frozen to preserve its learned diarization capabilities, while the remaining modules are fine-tuned to adapt to the spatial representations used for conditioning.
The system achieves a macro DER of \SI{11.7}{\percent}, corresponding to an improvement of 0.5 percentage points over the DiariZen baseline.
Notably, the improvement is consistent across all evaluated datasets.

To confirm that the observed improvements are not merely a result of increased parameter count, an additional experiment, DiariZen-Large Conformer (ID 2), was conducted in which the Conformer is doubled in size. As shown in \cref{tab:comparison}, this larger model does not lead to any performance gain, achieving a macro DER of \SI{12.9}{\percent}, indicating that the improvements achieved by DiariZen + Spatial Diarization (ID 7) can indeed be attributed to the integration of spatial cues rather than to increased model size.

Finally, the DiariZen + Spatial Diarization configuration with pretraining is fine-tuned (ID 8) with the spatial diarization auxiliary network unfrozen. This allows the spatial diarization module to adapt jointly with the Conformer and the modulation through the FiLM layers, resulting in a macro DER of \SI{11.6}{\percent}, an improvement of 0.6 percentage points over the purely spectral DiariZen system.

\subsection{Analysis and Discussion} %
\begin{table}[!tb]
    \centering
    \vspace{-6pt}
    \caption{Macro-averaged DER performance of selected systems using VBx clustering.}
    \label{tab:vbx}
    \vspace{3pt}
    \begin{tabular}{clc}
        \toprule 
        \textbf{ID} & \textbf{System} & \makecell{\textbf{Macro DER}\\(OV / Single)} \\
        \midrule
        {1} & DiariZen~\cite{han2025leveraging} & 14.8 (27.1 /\hphantom{1} 9.6)  \\
        \hline
        {4} & Spatial Diarization  &  16.8 (27.7 / 11.6) \\
        {7} & \diarizen{} + Spatial Diar.  & 14.3 (26.9 / \hphantom{1}9.2) \\
        {8} & \quad + Joint Finetuning &  14.1 (26.1 /\hphantom{1} 9.1) \\
        \midrule
        \textcolor{gray!90}{9} & \textcolor{gray!90}{DiariZen + Oracle \#Spk} & \textcolor{gray!90}{\hphantom{1}9.0 (22.4 / \hphantom{1}3.1)}\\
        \bottomrule
    \end{tabular}
\end{table}

As the results show, the spatially supported DiariZen system can improve the diarization performance compared to the single-channel system in an oracle clustering setting.
\cref{tab:vbx} further demonstrates that the observed improvements persist when employing VBx clustering instead of oracle clustering, as in ~\cite{han2025efficient}.
Similar relative gains over the baseline are achieved and indicate that the benefits achieved at the local EEND module level effectively transfer to the full diarization pipeline.
However, since the powerset output of the systems is restricted to a maximum of two concurrent speakers, performance is inherently limited in regions with three or more active speakers, which account for approximately \SI{3.5}{\percent} of the total duration in the evaluation data.

Overall, the inclusion of spatial information does not yield as large an improvement as initially expected for the DiariZen architecture. 
In particular, the single-channel system already performed surprisingly good in overlapping speech regions. 
A possible explanation, as discussed in ~\cref{cha:baselines}, is that the learnable weighted combination of all WavLM layers already gives the model access to information that helps to distinguish overlapping speakers.

An alternative approach to taking advantage of multi-channel input is to make WavLM multi-channel aware, as proposed in \cite{han2025spatial}. \mbox{ID 3} in \cref{tab:comparison} shows the achieved results (taken from \cite{han2025spatial}).
It can be observed that similar gains are obtained as with the method proposed here. We conclude that these are two concurrent approaches to making DiariZen multi-channel aware,  ours incorporating explicit spatial cues, and the other adapting the foundation model itself for multi-channel processing.

\section{Conclusions}
\label{sec:conclusions}

This work investigates whether and how spatial information can further improve a state-of-the-art single-channel diarization system based on self-supervised foundation model features. 
To this end, multiple strategies to incorporate spatial cues from multi-channel recordings into the DiariZen framework were analyzed.
While integrating the spatial cues using an untrained auxiliary encoder does not improve diarization performance, employing a spatial diarization network, pretrained for a diarization objective, leads to small but consistent gains across all evaluated datasets. 
The results confirm that spatial information can complement single-channel representations in realistic meeting scenarios. 
However, the gains are not concentrated in overlapping speech regions as initially expected.

\section{Acknowledgements}
\label{sec:prior}

The work reported here was started
during JSALT 2025 and supported by JHU. BUT researchers were supported 
by Ministry of Education, Youth and Sports of the Czech Republic (MoE) 
through the OP JAK project \mbox{CZ.02.01.01\/00\/23\_020\/0008518}. Computing on IT4I supercomputer was supported by MoE through the {e-INFRA CZ (ID:90254)}.

\bibliographystyle{IEEEbib}
\bibliography{refs}

\end{document}